\documentclass[sigconf]{acmart}

\usepackage{graphicx}
\usepackage{textcomp}
\usepackage{xcolor}
\usepackage{booktabs} 
\usepackage{listings}
\usepackage{subcaption}
\usepackage{multirow}
\usepackage{balance}
\usepackage{fixltx2e}
\usepackage{url}
\usepackage[multiple]{footmisc}
\usepackage[color=white]{todonotes}
\usepackage[export]{adjustbox}
\usepackage{xspace}
\usepackage{tcolorbox}
\usepackage{tabularx}
\usepackage{diagbox}
\usepackage{booktabs}
\usepackage{tikz}
\usepackage{multirow} 
\usetikzlibrary{shapes.geometric, arrows, positioning}
\usepackage{caption}
\usepackage{subcaption}
\usepackage{hyperref}
\usepackage{algorithm}
\usepackage{algpseudocode}
\usepackage{placeins}
\tikzstyle{startstop} = [rectangle, rounded corners, minimum width=3cm, minimum height=1cm,text centered, draw=black, fill=red!30]
\tikzstyle{io} = [trapezium, trapezium stretches body, trapezium left angle=70, trapezium right angle=110, minimum width=3cm, minimum height=1cm, text centered, draw=black, fill=blue!30]
\tikzstyle{process} = [rectangle, minimum width=3cm, minimum height=1cm, text centered, draw=black, fill=orange!30]
\tikzstyle{decision} = [diamond, minimum width=3cm, minimum height=1cm, text centered, draw=black, fill=green!30]
\tikzstyle{arrow} = [thick,->,>=stealth]

\copyrightyear{2024}
\acmYear{2024}
\setcopyright{rightsretained}
\acmConference[ICPE '24]{Proceedings of the 15th ACM/SPEC International Conference on Performance Engineering}{May 7--11, 2024}{London, United Kingdom}
\acmBooktitle{Proceedings of the 15th ACM/SPEC International Conference on Performance Engineering (ICPE '24), May 7--11, 2024, London, United Kingdom}
\acmDOI{10.1145/3629526.3645037}
\acmISBN{979-8-4007-0444-4/24/05}

\makeatletter
\newcommand*{\rom}[1]{\expandafter\@slowromancap\romannumeral #1@}
\makeatother

\begin{document}


\title{A Learning-Based Caching Mechanism for Edge Content Delivery}

\author{Hoda Torabi}
\orcid{0009-0003-0150-7321}
\affiliation{%
\institution{York University}
\department{Department of Electrical Engineering \& Computer Science}
\city{Toronto}
\country{Canada}}
\email{hodat@yorku.ca}
\author{Hamzeh Khazaei}
\orcid{0000-0001-5439-8024}
\affiliation{%
\institution{York University}
\department{Department of Electrical Engineering \& Computer Science}
\city{Toronto}
\country{Canada}}
\email{hkh@yorku.ca}
\author{Marin Litoiu}
\orcid{0000-0003-0383-920X}
\affiliation{%
\institution{York University}
\department{Department of Electrical Engineering \& Computer Science}
\city{Toronto}
\country{Canada}}
\email{mlitoiu@yorku.ca}

\begin{abstract}
With the advent of 5G networks and the rise of the Internet of Things (IoT), Content Delivery Networks (CDNs) are increasingly extending into the network edge. This shift introduces unique challenges, particularly due to the limited cache storage and the diverse request patterns at the edge. These edge environments can host traffic classes characterized by varied object-size distributions and object-access patterns. Such complexity makes it difficult for traditional caching strategies, which often rely on metrics like request frequency or time intervals, to be effective. Despite these complexities, the optimization of edge caching is crucial. Improved byte hit rates at the edge not only alleviate the load on the network backbone but also minimize operational costs and expedite content delivery to end-users.
In this paper, we introduce HR-Cache, a comprehensive learning-based caching framework grounded in the principles of Hazard Rate (HR) ordering, a rule originally formulated to compute an upper bound on cache performance. HR-Cache leverages this rule to guide future object eviction decisions. It employs a lightweight machine learning model to learn from caching decisions made based on HR ordering, subsequently predicting the ``cache-friendliness'' of incoming requests. Objects deemed ``cache-averse'' are placed into cache as priority candidates for eviction. Through extensive experimentation, we demonstrate that HR-Cache not only consistently enhances byte hit rates compared to existing state-of-the-art methods but also achieves this with minimal prediction overhead.
Our experimental results, using three real-world traces and one synthetic trace, indicate that HR-Cache consistently achieves 2.2–14.6\% greater WAN traffic savings than LRU. It outperforms not only heuristic caching strategies but also the state-of-the-art learning-based algorithm.
\end{abstract}



\maketitle

\section{Introduction} \label{introduction}
The increasing use of multimedia content services and the expansive deployment of IoT systems necessitate fast and dependable content delivery. Essential online services, such as web hosting and video streaming, rely heavily on the efficiency of these deliveries. With the advancement of 5G networks, edge caching has become a pivotal technology for addressing these rising demands and improving overall performance. By storing content closer to users at the network edge, edge caching plays an essential role in enhancing user experience and reducing the bandwidth required across the wide-area network between edge nodes and the original content servers.

This surge in demand for content delivery directly leads to a substantial and growing volume of traffic. This increase has a significant financial impact on network service providers, particularly if traffic is not managed effectively at the edge. While low latency is critical for delivering small-size, latency-sensitive content, the primary objectives for distributing larger files, such as video streams and extensive downloads, are to minimize traffic handling costs and prevent overload at network bottlenecks \cite{mokhtarian2014caching}. Therefore, a key goal of caching at the edge of network is to maximize the fraction of bytes served locally from the cache \cite{hasan2014trade}, also known as the byte hit ratio (BHR).

The caching algorithm, which decides which objects are cached, is integral to achieving a low byte miss ratio. As such, this problem has been extensively studied since the advent of the internet. Caching strategies have evolved from basic heuristic methods like Least Recently Used (LRU), which evicts the oldest data first, to intricate algorithms that combine frequency and recency (e.g. Hyperbolic) and others that use a composition of frequency and object size (e.g. GDSF). Despite extensive research, most production systems—such as those employed by Akamai \cite{nygren2010akamai}, Memcached, and NGINX—commonly implement LRU variants as their standard caching algorithm. Yet, these may not be ideally suited to the particular demands of edge caching, which contends with limited cache sizes and the unpredictable nature of user requests \cite{liu2016caching}. The challenge in designing effective caching algorithms is that workload characteristics, like object access patterns or request processes, are not constant and often change over time. Thus, a heuristic that performs well for one workload scenario may falter in another, or fail to adapt when access patterns evolve, underscoring the necessity for flexible caching strategies that can overcome these challenges.

Recent advancements in machine learning (ML) have opened new avenues for enhancing cache algorithms, particularly in the face of the aforementioned challenges. One increasingly popular method is to employ ML techniques to forecast the popularity of objects for proactive caching \cite{shuja2021applying}. Yet, this method can lead to suboptimal cache performance \cite{ferragut2016optimizing} under some workloads and proactive downloads consume additional bandwidth which is counter-intuitive to our optimization goal. Another popular approach employs model-free reinforcement learning (RL), where the system starts with no preconceived notions about the traffic patterns and iteratively works its way toward the optimal caching strategy \cite{kirilin2019rl, zhong2018deep, sadeghi2019deep, wang2020federated}. In this system, decision-making is honed through direct interaction with the environment, utilizing a reward function to reinforce actions that yield positive outcomes. Due to the large state-action space, RL methods tend to be more complex and require greater computational resources. Additionally, they can be sensitive to hyper-parameters, making it challenging to fine-tune their performance. 

A promising approach for designing cache algorithms has been to leverage oracle policies, such as the offline Belady algorithm \cite{belady1966study}, the practical flow-based offline optimal (PFOO) algorithm \cite{berger2018practical}, and the more recent Hazard Rate (HR) based upper bound \cite{panigrahy2022new}, which compute the theoretically optimal cache decisions. This is achieved by either learning to ``imitate'' the optimal decisions \cite{liu2020imitation, berger2018towards} or by directly learning and predicting the objects’ next request arrival to inform the optimal caching choice as explored by \cite{song2020learning}.
In this paper, we introduce HR-Cache framework, a new learning-based caching framework grounded in the principles of Hazard Rate Ordering~(HRO) rule introduced in \cite{aho1971principles}. HR-Cache is based on several original contributions. The framework is divided into two main components: The first component calculates the caching decisions for a window of past requests based on the HRO rule. The second part then trains an ML model that maps a set of features to HRO cache decisions. This model is then applied to in the next window to predict the ``cache-friendliness'' of objects at the time they are requested. In the event of a cache miss, where eviction is necessary to make space, our framework preferentially evicts items in the cache that were previously identified as ``cache-averse.''

Our application of the HRO rule presents an intricate challenge: accurately determining the hazard rate function for object inter-request times to reconstruct the HRO, without making simplifying assumptions about the nature of the request distribution. We address this by employing a Kernel Hazard Estimator, which estimates the hazard function directly from the data without assuming a specific parametric form for the distribution. This consideration can be particularly important, as our use of ML methods is intended to address the shortcomings of heuristic-based algorithms, which usually excel with specific access patterns but not others. Therefore, making assumptions about the workload might negate the advantages that machine learning brings to our caching decision process. Putting all this together in a practical system, however, requires us to address other challenges including controlling the computational overhead for ML training and prediction.
Our evaluation results using production and synthetic traces show that our learning-based policy consistently performs better than state-of-the-art methods and reduces WAN traffic by 4–25\% compared to the LRU replacement policy and reduces the prediction overhead by a factor of 19.2x compared to the state-of-the-art learning-based cache policy.
\textbf{Roadmap:} We organize the rest of this paper as follows. Section 2 provides background and discusses related work. Section 3
presents the HR-Cache algorithm. Section 4 empirically evaluates
the proposed scheme. Section 5 concludes the paper with a summary of its contributions.

\section{Background \& Related Work} \label{background}
Most earlier designs of caching rely on heuristic-based methods including the least recently used (LRU), least frequently used (LFU), and first in first out (FIFO), along with their variants.  While these classical methods offer straightforward solutions for managing cache resources, they often fall short in adapting to the dynamic and complex nature of request patterns. Moreover, different features may have varying levels of importance across diverse workloads, a nuance that heuristic methods struggle to accommodate. This limitation is particularly evident in edge networking environments, where traditional traffic assumptions may no longer be valid \cite{traverso2015unravelling, paschos2016wireless}. In addition, prior work \cite{song2020learning, berger2018towards} have pointed out a considerable discrepancy between current state-of-the-art caching designs and theoretical upper limits on cache performance, as established by algorithms like Belady's algorithm \cite{belady1966study}, flow-based offline optimal \cite{berger2018practical}, and the hazard rate upper bound \cite{panigrahy2022new}. This significant gap, along with the supporting evidence from recent measurement studies in edge caching systems \cite{huang2013analysis, song2017learning, guan2019caca}, signals a clear opportunity for enhancements in cache performance and addressing the limitations of existing caching strategies. In light of these developments, there has been a growing emphasis in recent research on developing learning techniques that can intelligently manage cache resources. In the following part, we will focus on a discussion of learning-based cache policies, providing the necessary background and rationale for the development of the proposed HR-Cache framework. Simultaneously, this portion will serve as a review of related work in this area. We'll finish this section with a brief overview of use of ML methods to improve performance, bringing together the key aspects of our research discussion.

\subsection{Learning-based Caching}
Approaches to learning-based caching can be roughly grouped into three categories, with the first category encompassing recent research efforts focused on leveraging theoretically optimal caching policies to develop learning-based methods. A significant point of reference here is the Belady optimal policy \cite{belady1966study}. This algorithm operates on the principle of evicting the object that will be used furthest in the future, thereby minimizing miss rate. While Belady's algorithm provides an ideal strategy for cache replacement, its real-world application has been limited because it requires foreknowledge of future cache access patterns, which is generally not feasible. 
Nevertheless, this algorithm forms a basis for designing practical caching policies. 

Hawkeye \cite{jain2016back} was the first to introduce learning from the Belady’s algorithm.  Hawkeye employs a binary classification model to determine whether a cache line is likely to be reused (deemed ``cache-friendly'') or not (``cache-averse''). Their policy prioritizes the eviction of cache-averse lines over cache-friendly ones. By using oracle labels for previous access patterns, Hawkeye effectively transforms cache replacement into a supervised learning challenge. Building upon Hawkeye's foundation, Glider \cite{shi2019applying} enhances this approach by integrating deep learning techniques to develop a more accurate predictor than its predecessor. However, it’s important to note that both Hawkeye and Glider focus on hardware caches and are not directly applicable to software cache systems, particularly those handling variable-sized objects. Another work, ``Parrot'' as described in \cite{liu2020imitation} adopts an imitation learning approach to automatically learn cache access patterns by leveraging Belady’s. Although effective, its computational demands can be significantly high.

Diverging from Parrot’s methodology, LRB, as outlined in \cite{song2020learning}, employs a different strategy by predicting the next arrival times of object requests. This enables LRB to approximate Belady's algorithm through a supervised learning method. By learning the next access time for each object based on a multitude of features, LRB identifies and evicts objects predicted to have the furthest request time. This strategy has demonstrated enhanced performance over state-of-the-art caching algorithms in terms of byte hit ratios.
However, LRB is not without its limitations. To closely emulate the optimal offline oracle, a system like LRB is required to predict the next access times for all objects in the cache, selecting for eviction the one with the most distant future request. This prediction process can be extremely resource-intensive for large caches. LRB mitigates this by limiting the inference to a sample of 64 objects for each eviction. Despite this optimization, the prediction overhead remains a significant computational burden. LRB's use of dynamic features means that prediction results are not reusable over time, necessitating fresh sampling and inference for every eviction. Reflecting this overhead, LRB's simulation shows that on a single CPU core, each eviction in LRB consumes 227.19 µs.\footnote{For 64 GB cache size, Wikipedia 2019 workload} Consequently, this caps the eviction rate at a maximum of approximately 4,500 objects per second per core, rendering it less efficient for high-demand production environments.

LFO \cite{berger2018towards}, another work employing supervised learning, first calculates the sequence of optimal caching decisions (OPT) for recent history using a min-cost flow model from \cite{berger2018practical}, designed for optimal caching of variable-sized objects. Following this calculation, LFO applies manually-designed features and a gradient boosting decision tree to train a binary classifier for caching decisions. The classifier's prediction is then used to imitate the admission policy of OPT and serve as a ranking metric for the eviction policy. However, the process of deriving optimal decisions based on the min-cost flow model is complex and computationally intensive, hindering LFO's ability to swiftly adapt to workload changes. Additionally, its design necessitates executing a prediction for every incoming request, further impacting its practical efficiency.

Inspired by similar principles to our work, LHR in \cite{yan2021learning} draws on the concept of the Hazard Rate bound from \cite{panigrahy2022new} to develop a learning-based caching policy. Unlike a direct adoption, LHR modifies this approach by constructing an online upper bound, which approximates the request process through a Poisson process. Under this assumption, the hazard rate remains constant and is equivalent to the request rate for each object. While this approach simplifies their model, it considerably narrows the applicability of LHR \cite{hu2022raven}; particularly in light of \cite{panigrahy2022new}'s demonstration that the HRO upper bound is effective for any stationary arrival process. Thus, LHR's reliance on the Poisson assumption potentially restricts the full exploitation of HRO's capabilities.

Considering the insights gained from the review of existing works, our framework's design will be informed around these pivotal lessons:
\begin{enumerate}
    \item \textbf{Utilization of HRO Bound:} Taking into account the limitations of LHR's Poisson assumption, our approach will seek to fully leverage the HRO bound's potential, avoiding oversimplified assumptions that could undermine the practicality and justification of using machine learning.
    \item \textbf{Minimizing Prediction Overhead:} Addressing the challenge of high computational demands seen in methods like LRB, our framework will prioritize efficient prediction mechanisms to enhance scalability and performance.
    \item \textbf{Decision-Making Process:} Considering the complexity of the LFO approach, we aim to create an efficient method for making caching decisions. This is important in fast-paced environments where models need regular updates and training. Our approach is designed for quick adjustments to stay up-to-date with frequent changes.
\end{enumerate}

\subsection{Use of Machine Learning to Improve Performance}
Our work is part of the growing effort to use machine learning to improve system performance, especially in caching strategies. This effort extends across various fields, as evidenced by related literature showcasing the application of machine learning (ML) and deep learning (DL) in enhancing system performance and efficiency in different domains. For instance, the work in \cite{tuli2022optimizing} employs deep neural networks (DNNs) for optimizing resource management in edge computing environments, enabling dynamic scheduling in distributed fog systems by estimating key Quality of Service (QoS) metrics. The work of \cite{garbi2020learning} utilizes recurrent neural networks to develop performance models for queuing networks, aiming to improve resource utilization based on it. In the realm of database efficiency, the work in \cite{marcus2020bao} leverages tree convolutional neural networks and reinforcement learning to optimize queries , while the work in \cite{nathan2020learning} applies machine learning techniques for effective database indexing. Collectively, these studies underscore the potential of ML and DL as powerful tools for system performance optimization.

\section{HR-Cache} \label{motivational}
In this section, we begin by discussing the hazard rate upper bound introduced in \cite{panigrahy2022new}, which forms the cornerstone of our approach. Building upon this foundation, we then introduce our learning-based caching policy, HR-Cache. The primary goal of HR-Cache is to assess whether a requested object is cache-friendly or cache-averse. Upon a cache miss, the requested object is inserted into the cache; however, objects identified as cache-averse are placed in a candidate queue for potential future eviction. HR-Cache gives priority to evicting objects from this candidate queue, resorting to the main queue only when the candidate queue becomes empty.

\begin{figure}[t]
  \centering
    \includegraphics[width=0.95\linewidth]{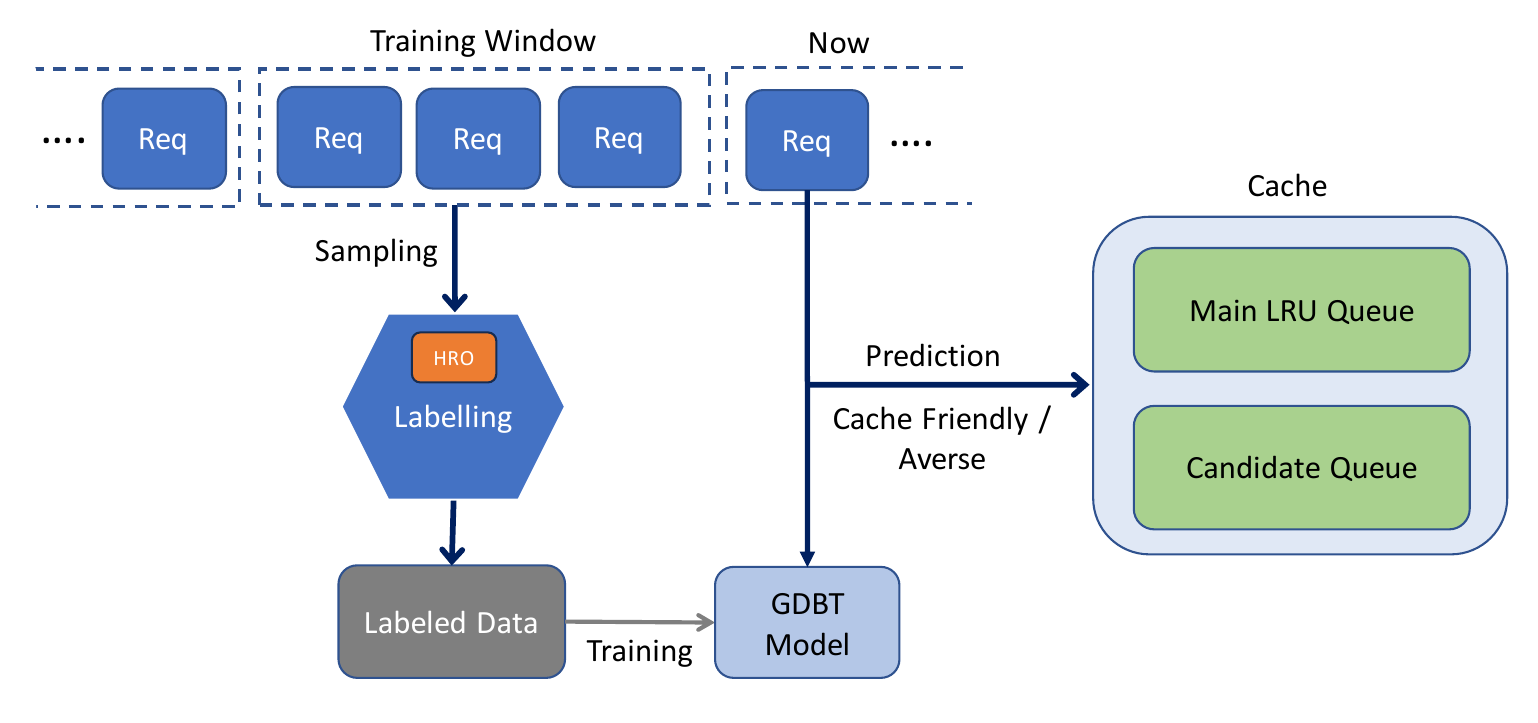}
    \caption{Architecture Overview of HR-Cache}
  \label{fig:hrcache_arch}
\end{figure}

To learn how to make this decision, our method reconstructs the hazard rate ordering solution from \cite{panigrahy2021resource} for past requests within a window to provide training targets for our model. The output from the model is then used in our caching policy. 

Figure \ref{fig:hrcache_arch} provides a high-level overview of the HR-Cache architecture.

\subsection{Hazard Rate-Based Upper Bound}\label{sec: hro_define}
The framework in \cite{panigrahy2022new} considers a caching system serving $n$ distinct objects, possibly of different sizes. The system is defined with a cache capacity of $B$ bytes, where $B$ is a predefined parameter representing the total storage capacity of the cache. In the basic case, the model assumes a cache of size 
$B$ bytes addressing requests for $n$ distinct objects of equal size. In this context, \cite{panigrahy2022new} introduces the hazard rate based rule, termed as HR-E, which operates as follows:
\begin{itemize}
\item At any given time $t$, HR-E first determines the hazard rate function for each object.
\item Then it places in the cache the $B$ objects which have the largest hazard rates (ties between equal rates are broken randomly) .
\item A request at time $t$ is considered a ``hit'' if the requested object is among those cached based on the aforementioned criteria.
\end{itemize}
They use this rule as a way to upper-bound various cache performance metrics including object hit and byte hit ratio.

They further extend this rule to obtain an upper bound on the byte hit probability for variable size objects. In this case, the authors adapt the hazard rate-based rule denoted as HR-FC to accommodate fractional caching, a strategy that permits the storage of a fraction of an object. Specifically, the rule at any time caches objects with the highest hazard rates until an object cannot fit. For the object that cannot be fully fit due to limited remaining cache capacity, only a sufficient number of bytes required to reach the cache limit are stored.
In the case for equal-sized objects, the HR-E rule serves as an upper bound on the cache hit probability for non-anticipative caching policies, while HR-FC serves as an upper bound on the cache byte hit probability, which is the metric we are interested in. Throughout this work, we will collectively refer to these rules as the ``HRO'' rule (hazard rate ordering rule) for consistency and ease of reference.

\subsection{Hazard Rate Function}
Let us consider the sequential times at which object $i$ is accessed as $\{\tau_{ik} \mid k \in \mathbb{Z}\}$. The time interval between consecutive requests---namely, the $k$th and $(k-1)$th requests---for the same object $i$ is termed $X_{ik}$ and computed as $\tau_{ik} - \tau_{i(k-1)}$, for $k \geq 1$. By default, $\tau_{i0}$ is set to zero. The sequence $\{X_{ik}\}_{k \geq 1}$ is assumed to form a stationary point process, with the cumulative distribution function (c.d.f) for the inter-arrival time given as $F_i(t) = \text{P}(X_{ik} \leq t)$, and its corresponding density function is represented as $f_i(t)$.

The hazard rate function, denoted as $\lambda_i(t)$, associated with $F_i(t)$ is defined as follows:
\begin{equation}
\lambda_i(t) = \frac{f_i(t)}{1 - F_i(t)}, \quad t \in [0, F_i^{-1}(1)],
\end{equation}

Here, the hazard rate function is the conditional density of the occurrence of an object request, given the realization of the request process over $[0, t)$ \cite{daley2014introduction}.
It is noteworthy that the hazard rate function's meaning can vary based on its application context. For example, in survival analysis, the hazard rate quantifies the conditional probability of an item's failure, given that it has remained functional up to a specific time point. In caching terminology, failure of an object can be treated as the object being requested.

\subsection{Calculating Hazard Rate}\label{sec:calcHR}
To effectively implement hazard rate-based rule in our framework, we must first accurately determine the hazard rate function for each object. While this is relatively straightforward for synthetic data sets, it poses a significant challenge in real-world production settings. One approach to this challenge is approximating the inter-request times of objects using well-defined distributions, such as Poisson \cite{yan2021learning} or Generalized Pareto \cite{panigrahy2022new}. However, relying solely on these approximations could potentially diminish the benefits of leveraging machine learning in cache decision-making since these approximations may not be universally applicable across varying workloads and use-cases.
Therefore, to calculate hazard rates that are adaptable to various workload trace distributions, we use the kernel hazard estimator proposed by \cite{muller1994hazard}. We obtain this estimator by applying smoothing to the increments of the Nelson-Aalen estimator. 

The Nelson-Aalen estimator is a non-parametric method used to estimate the cumulative hazard function in survival analysis. Unlike parametric methods, which make specific assumptions about the underlying hazard rate distribution, the Nelson-Aalen estimator does not require any such assumptions. 
We denote $H(t)$ as the cumulative hazard function at time t. 
The estimator is given by:
\[
H(t) = \sum_{j: t_j \leq t} \frac{d_j}{n_j}
\]
\\where \( t_j \) are the observed event times, \( d_j \) is the number of events at time \( t_j \), and \( n_j \) is the number of subjects at risk just before time \( t_j \).
However, the Nelson-Aalen estimator results in a step function, which is not differentiable. Instead, kernel smoothing techniques are utilized to smooth the increments of the cumulative function estimate obtained by the Nelson-Aalen estimator \cite{wang2005smoothing}. The kernel hazard estimator we use takes the form:

\[
\lambda(t) = \frac{1}{h} \sum_{i=1}^{n} K\left(\frac{t-t_i}{h}\right) \Delta H(t_i)
\]
where \( K(\cdot) \) is a kernel function (e.g., Epanechnikov kernel), \( h \) is the bandwidth, determining the width of the smoothing window, and \( \Delta H(t_i) \) is the increment in the Nelson-Aalen estimate at time \( t_i \), which is \( \frac{d_i}{n_i} \).

\begin{table}[h]
\centering
\begin{tabular}{|c|c|}
\hline
Trace length & 3.7 million \\ \hline
Unique objects & 5638 \\ \hline
\end{tabular}
\vspace{3pt}
\captionsetup{width=0.95\linewidth}
\caption{IBM Web Access Trace Collected from a Gateway Router}
\vspace{-8mm} 
\label{table:ibm_trace}
\end{table}

\begin{figure}[htbp]
  \centering
    \includegraphics[width=0.95\linewidth]{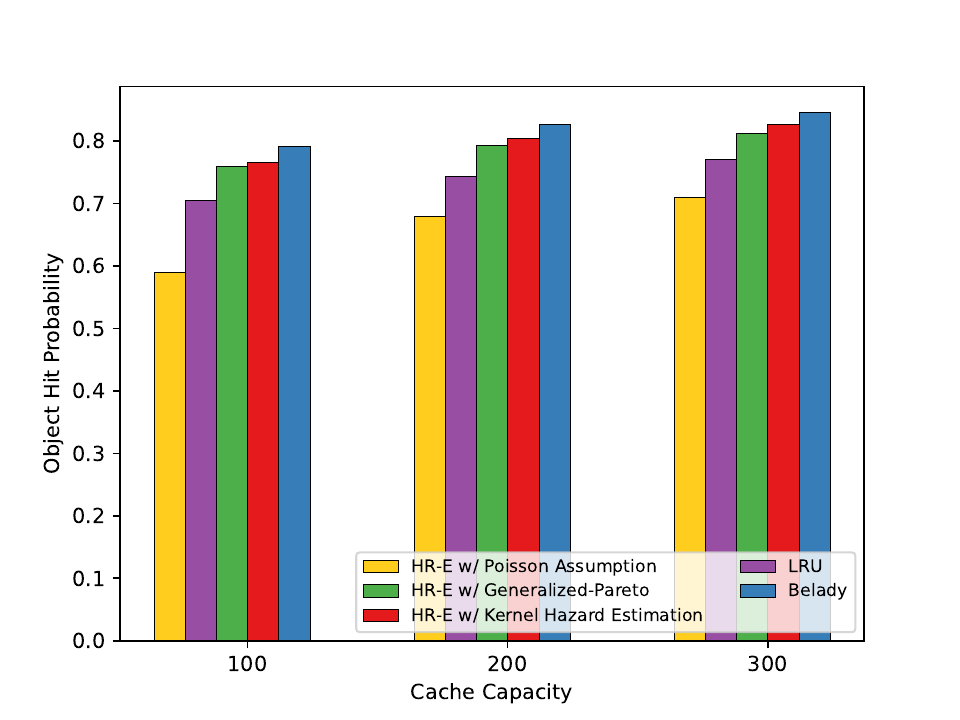}
    \caption{HR-E Upper Bound Comparisons and Hit Probabilities for LRU and Belady's Algorithms Across Three Cache Sizes.}
  \label{fig:hre_bound_fig}
\end{figure}

We use a real-world IBM trace from \cite{panigrahy2022new}, to test the validity of the non-parametric hazard estimator. Details about the trace are provided in Table \ref{table:ibm_trace}.
For our experiment, we derive the upper bound on hit probability using the HR-E ordering rule with three different estimators. The first method employs the non-parametric estimator introduced earlier in this section. The study by \cite{panigrahy2022new} effectively estimated the hazard rate for each object in the IBM trace, assuming a Generalized Pareto distribution for inter-request times. We include the HR-E upper bound calculated under their estimator for validation. Additionally, we explore the HRE upper bound assuming request processes follow a Poisson process. For further comparison, we also present the object hit probabilities attained by the LRU and Belady's algorithms. The results of these comparisons, run under 3 different cache sizes, are illustrated in Figure \ref{fig:hre_bound_fig}.

As can be seen, the kernel hazard estimation method we use gives us an upper bound that aligns with the expected bound derived using the ``good'' parametric estimator of the Generalized Pareto distribution. This confirms that kernel hazard estimation is indeed suitable for our use case. As anticipated, the simplistic nature of the Poisson assumption results in LRU outperforming it. Moreover, our results reaffirm the tighter upper bound achieved by the HR-E rule compared to the Belady algorithm, consistent with the findings of the work in \cite{panigrahy2022new}.

\subsection{Learning From HRO}\label{sec: derive_hrrule}
Before diving into the learning process, it is crucial to make a key observation. We argue that since the HRO bound (Section \ref{sec: hro_define}) is derived in a pre-fetching manner, it does not directly correspond to cache decision at the time of request to an object. Specifically, the HRO rule assumes that at any time $t$, the objects with the highest hazard rates among all available objects, have already been pre-fetched and are present in the cache. Thus, the requested object at time $t$ is considered a hit if it is among objects in the cache. Based on this, for object $i$ to be considered as cached in the system, the previous request to object $i$ must admit it to the cache. Or in other words, when the request at time $t$ arrives, it can only be considered a hit if object $i$ was already cached due to a prior request. We classify these earlier requests as cache-friendly, as they are the ones leading to hits. With this and the HRO rule as a backdrop, we are set to develop a learning-based caching strategy. Our approach employs a sliding window of past requests $W[k]$. Using the gathered requests in $W[k]$, we do three things: 
\begin{enumerate}
\item First, using the inter-request times of objects in the window we calculate the increments of hazard rates for each object according to the Nelson-Aalen estimator, to be later smoothed by the kernel hazard estimator.
\item Second, we go over the requests in the window and mark them as hit/miss based on the HRO rule. Meaning for each request at time $t$, we compute the hazard rate at time $t$ for every object within the window using the kernel estimator, and consider objects with the highest hazard rates in cache until one doesn't fit. If the object requested at time $t$ is among the cached objects, it is considered as a hit; otherwise, it is considered as a miss.
\item Next, we examine the requests in the window once more: For each request $i$ that was marked as a hit in the first step, we mark the previous request to $i$ as cache-friendly. This provides us with a vector of cache decisions for requests in the window, which serves as the label data for our machine learning model training.
\end{enumerate}
HR-Cache then trains a model that maps features to the decision derived in step 3. The trained model is subsequently used over the next window, $W[k + 1]$, to inform cache decisions during which HR-Cache again records the requests to use for the next window and so on.

\subsection{Training Data}
An important design issue involves determining the optimal amount of past information to utilize. We adopt a sliding window approach, using the data within this window for hazard estimation, deriving the HRO cache decision, and model training. The choice of window size significantly impacts the system's effectiveness. A small window might result in few data for training or hazard rate estimations, while a window that is too large could lead to increased memory usage, as well as longer processing and training times.
While some studies arbitrarily define their window sizes (e.g., \cite{berger2018towards} opts for a window of 1 million, \cite{kirilin2019rl} for the initial 10 million requests), \cite{jain2016back} considers window size as a factor of cache capacity, $1\times$ represents a window that consists of accesses to k cache lines, where k is the capacity of the cache. We choose a $3\times$ window, meaning the unique bytes of object requests in the window is three times the cache size as we find that this works well across all our experiments, however there is room for investigating how to set an optimal window size.
In practice, the sliding window can encompass millions of objects, which presents significant challenges for the labeling process, particularly when reconstructing the HRO-Rule. To address this issue, HR-Cache employs a strategy of randomly sampling objects within the window to generate training samples. The sampling rate is automatically calibrated to ensure that the total number of operations stays within a manageable range, thus preventing the computational overhead from becoming prohibitive. In our C++ implementation, this adaptive approach has proven to be effective, yielding favorable results while keeping the computational demands at a reasonable level.

\subsection{Features}
When designing a machine learning model for cache decision, it is essential to choose relevant features that can help predict the optimal decision. Our chosen features encompass both the insights from past heuristics and the insight of recent learning-based caching policies.
Traditional caching heuristics focus on individual metrics, such as object recency (as seen in LRU), its frequency (as in LFU), or object size. This is while learning-based methods allow us to incorporates a range of them. We consider the following features which can be derived in an online and robust manner.
\begin{enumerate}
\item Delta series: The time differences between consecutive requests for an object. $\Delta_1$ indicates the amount of time since an object was last requested. $\Delta_2$ indicates the time in between an object's previous two requests and so on, i.e., $\Delta_n$ is the amount of time between an object's $n^{\text{th}}$ and $(n-1)^{\text{th}}$ previous requests. This can provide insights into the object's access pattern, which can help predict future requests. We use 32 deltas as our features.
\item Decayed frequency: Unlike simple frequency, decayed frequency accounts for the recency of requests by giving more weight to recent accesses. It calculates the fraction of requests for an object among all requests so far, but with a diminishing emphasis on older requests. This approach helps in capturing not just how often an object is requested, but also how its popularity or relevance changes over time.
\item Static features. These include unchanging characteristics of an object, such as its size and type. Static features can be useful due to their inherent correlation with different access patterns. For our implementation we only consider size among static features due to the availability of data in our traces.
\end{enumerate}

\subsection{Training HR-Cache}
The goal of HR-Cache is to map its features to a decision of whether
an incoming line is cache-friendly or cache-averse according to the HRO rule. For this task, we employ gradient boosting decision tree (GBDT) model. GBDTs are known for their strong performance across various datasets, particularly useful in tasks involving classification and regression with structured data. They are also convenient because they don't need feature normalization. Additionally, their effectiveness in caching-specific tasks is supported by studies like \cite{song2020learning} and \cite{berger2018towards}.

\begin{algorithm}
\caption{HR-Cache Policy}
\label{alg:hrcache}
\begin{algorithmic}[1]
\Procedure{UpdateCache}{$object$, $lookupTable$}
    \State Perform lookup for $object$ in $lookupTable$
    \If{$object$ is in cache (Hit)}
        \If{$object$ is in Candidate Queue}
            \If{predicted as Cache-friendly}
                \State Change mode to Main Queue
                \State Move $object$ from Candidate to Main Queue
            \EndIf
        \ElsIf{$object$ is in Main Queue}
            \If{predicted as Cache-friendly}
                \State Promote $object$ to MRU in Main Queue
            \ElsIf{predicted as Cache-averse}
                \State Change mode to Candidate Queue
                \State Move $object$ to Candidate Queue
            \EndIf
        \EndIf
    \Else \Comment{Request not in cache (Miss)}
        \If{predicted as Cache-friendly}
            \State Add $object$ to Main Queue
        \Else
            \State Add $object$ to Candidate Queue
        \EndIf
    \EndIf
\EndProcedure

\end{algorithmic}
\end{algorithm}

\begin{figure}[htbp]
  \centering
    \includegraphics[width=0.95\linewidth]{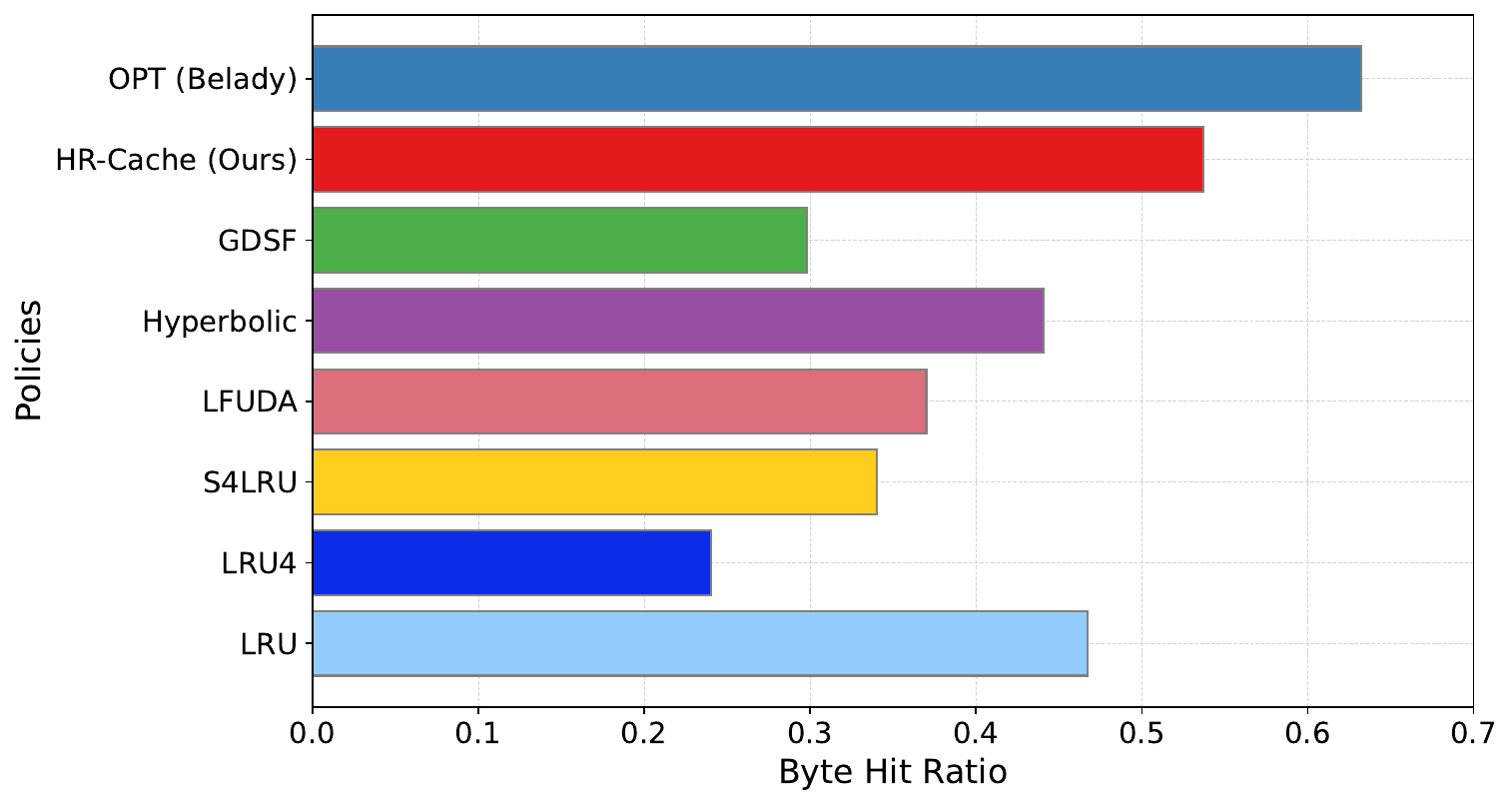}
    \caption{Comparison of HR-Cache to State-of-the-Art Heuristic Caching Systems for the IBM trace.}
  \label{fig:ibm_hrcache}
\end{figure}

\begin{table*}[h]
\centering
\caption{Summary of the traces used in our evaluation.}
\label{table:summary}
\begin{tabular}{|c|c|c|c|c|c|}
\hline 
\multicolumn{2}{|c|}{} & \textbf{Wikipedia 2018} & \textbf{Wikipedia 2019} & \textbf{CloudPhysics} & \textbf{EU Synthetic} \\
\hline 
\multicolumn{2}{|c|}{\textbf{Total Requests}} & 84 million & 90 million & 27 million  & 100 million \\
\hline 
\multicolumn{2}{|c|}{\textbf{Unique Objects Requested}} & 7 million & 11 million & 8 million & 41 million \\
\hline 
\multicolumn{2}{|c|}{\textbf{Total Bytes Requested}} & 2.6 TB & 3.4 TB & 360 GB & 100 TB \\
\hline 
\multicolumn{2}{|c|}{\textbf{Unique Bytes Requested}} & 0.75 TB & 1 TB & 86 GB & 38 TB \\
\hline 
\multirow{2}{*}{\textbf{Request Obj Size}} & \textbf{Mean} & 34 KB & 41 KB & 14 KB & 1 MB \\
\cline{2-6} 
& \textbf{Max} & 674 MB & 558 MB & 1 MB & 7 MB \\
\hline
\end{tabular}
\end{table*}

\subsection{The HR-Cache Policy}
Putting it all together, we design a caching policy guided by our learned model. For every object request, our HR-Cache predictor outputs a decision indicating whether the object is cache-friendly or cache-averse. This decision guides how we update the cache as detailed in Algorithm~\ref{alg:hrcache}. The goal is to manage objects so that cache-averse items end up in the candidate queue, while cache-friendly ones are placed in the main queue. The candidate queue consists of objects that the HR-Cache identifies as unlikely to lead to hits, hence prioritized for eviction. Meanwhile, the main queue employs an LRU strategy, ensuring that, should it become necessary to evict items from the main queue (once the candidate queue is depleted), the items least recently used are evicted first.

For a preliminary evaluation, we use the IBM request trace from Section~\ref{sec:calcHR} to assess the effectiveness of our learning framework. Given the trace's limited length, we use the initial one million requests to derive HRO decisions as outlined in Section~\ref{sec: derive_hrrule}. Subsequently, we train a model based on these decisions and apply the HR-Cache policy to evaluate the byte hit ratio on the remainder of the trace. As depicted in Figure~\ref{fig:ibm_hrcache}, HR-Cache demonstrates its effectiveness by achieving a Byte Hit Ratio that surpasses the state-of-the-art heuristic policies, even within the limited range of this relatively short trace.

\section{Experimental Evaluation} \label{setup}
We developed our framework in C++ as part of a trace-driven simulator designed to accurately assess our framework's miss ratios by replaying cache requests from traces. For the implementation of the Gradient Boosted Decision Trees (GBDT) model, we utilized the LightGBM framework \cite{ke2017lightgbm}. The code is publicly available on our GitHub repository\footnote{\url{https://github.com/pacslab/HR-Cache}} to facilitate the reproducibility of our proposed research in this paper. Additionally, we introduce an optimization in our implementation in the following section and examine its impact in Section \ref{sec:optimization}.

\subsection{Batched Predictions}
The basic HR-Cache needs to predict cache-friendliness of objects as each request arrives. To take advantage of the architectural strengths of multi-core processors in contemporary CDN and edge servers, we implement data parallelism in our cache decision-making. This modification permits parallel predictions for $N$ requests simultaneously. The chosen batch size, $N$, plays a critical role in balancing parallelism and miss ratio. A small $N$ fails to fully utilize the potential of parallelism, while an excessively large $N$ can lead to delayed predictions and negatively affect the miss ratio. We selected a batch size of $N$ = 128, finding it optimal for harnessing parallelism without affecting our miss ratio. For instance, in our simulations, which do not account for object retrieval overhead, a batch size of $N$ = 128 enabled an increase in throughput from handling 11,828 requests per second to 98,404 requests per second, while maintaining cache performance efficiency on the Wiki 2019 trace.

We conduct trace-driven simulations to evaluate the performance of HR-Cache against a broad spectrum of state-of-the-art caching algorithms. Our analysis primarily focuses on two key questions: First, we examine how the byte miss ratio of HR-Cache compares with that of other state-of-the-art research systems across a variety of traces and cache sizes. Second, we assess how HR-Cache performs in relation to the state-of-the-art (SOA) learning-based cache mechanisms, particularly in terms of prediction overhead.

\begin{figure*}[t!]
  \centering
  \begin{subfigure}{0.49\textwidth}
    \includegraphics[width=\linewidth]{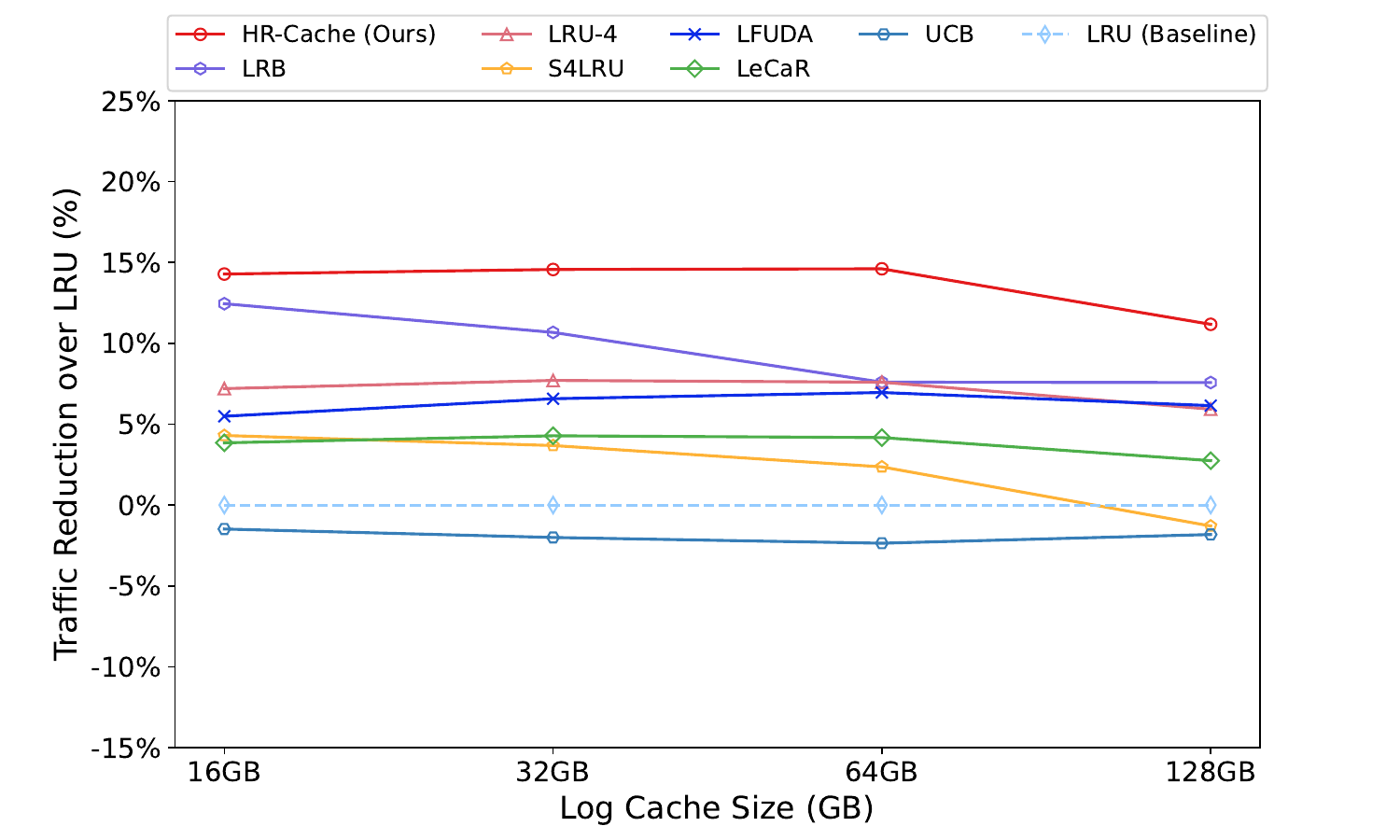}
    \caption{Wikipedia 2018}
    \label{fig:wiki18_traffic}
  \end{subfigure}
  \hfill
  \begin{subfigure}{0.49\textwidth}
    \includegraphics[width=\linewidth]{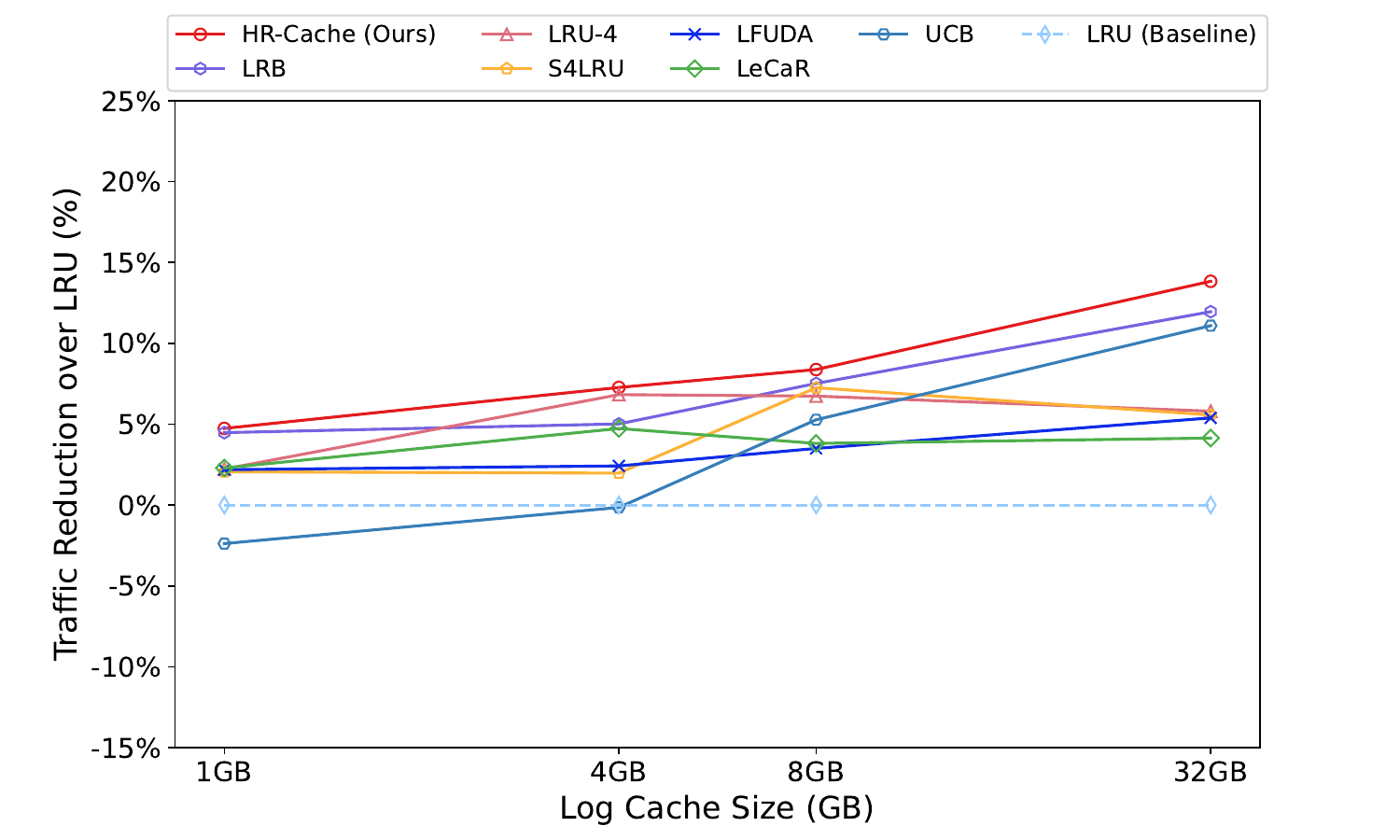}
    \caption{CloudPhysics}
    \label{fig:w16_traffic}
  \end{subfigure}

  \begin{subfigure}{0.49\textwidth}
    \includegraphics[width=\linewidth]{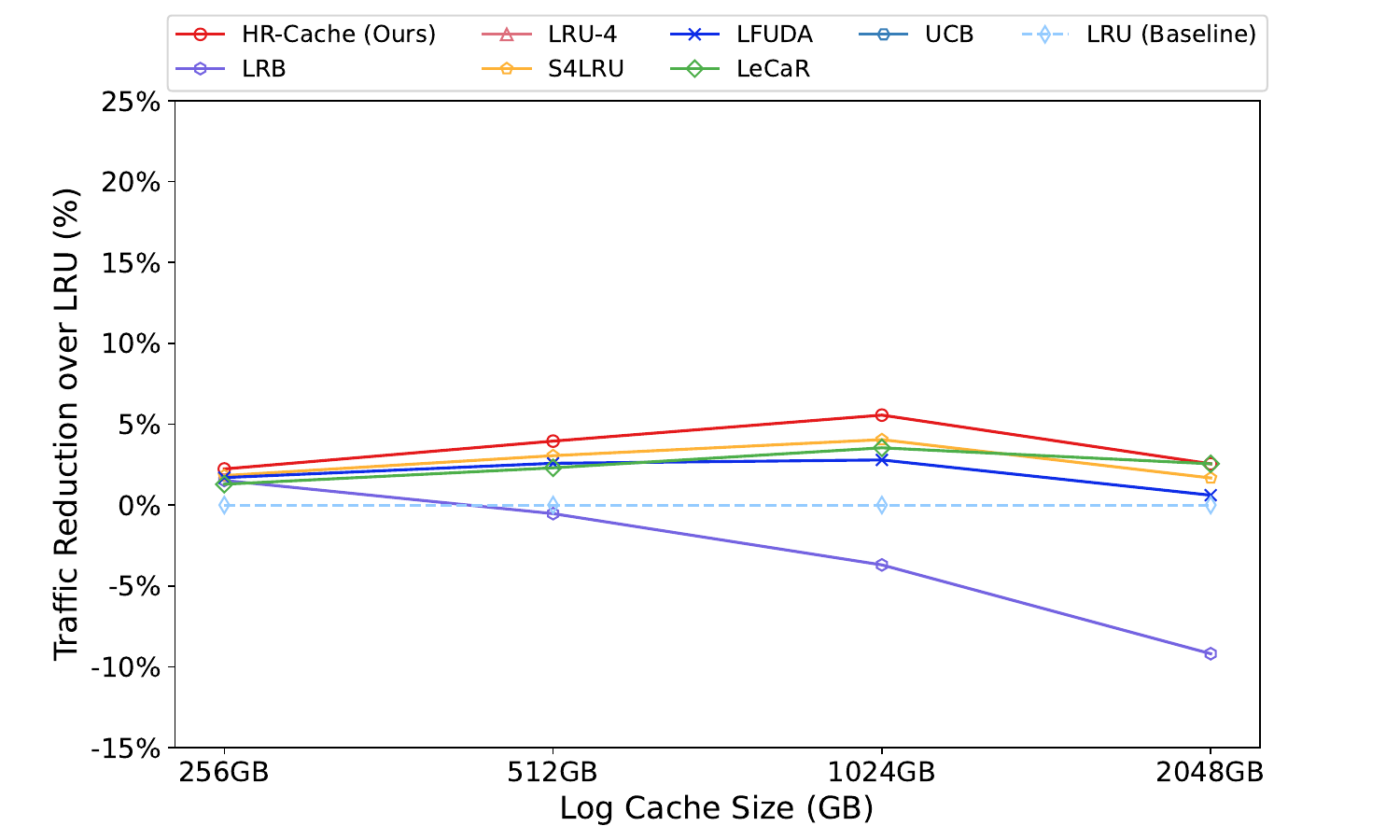}
    \caption{EU Synthetic}
    \label{fig:eu_traffic}
  \end{subfigure}
  \hfill
  \begin{subfigure}{0.49\textwidth}
    \includegraphics[width=\linewidth]{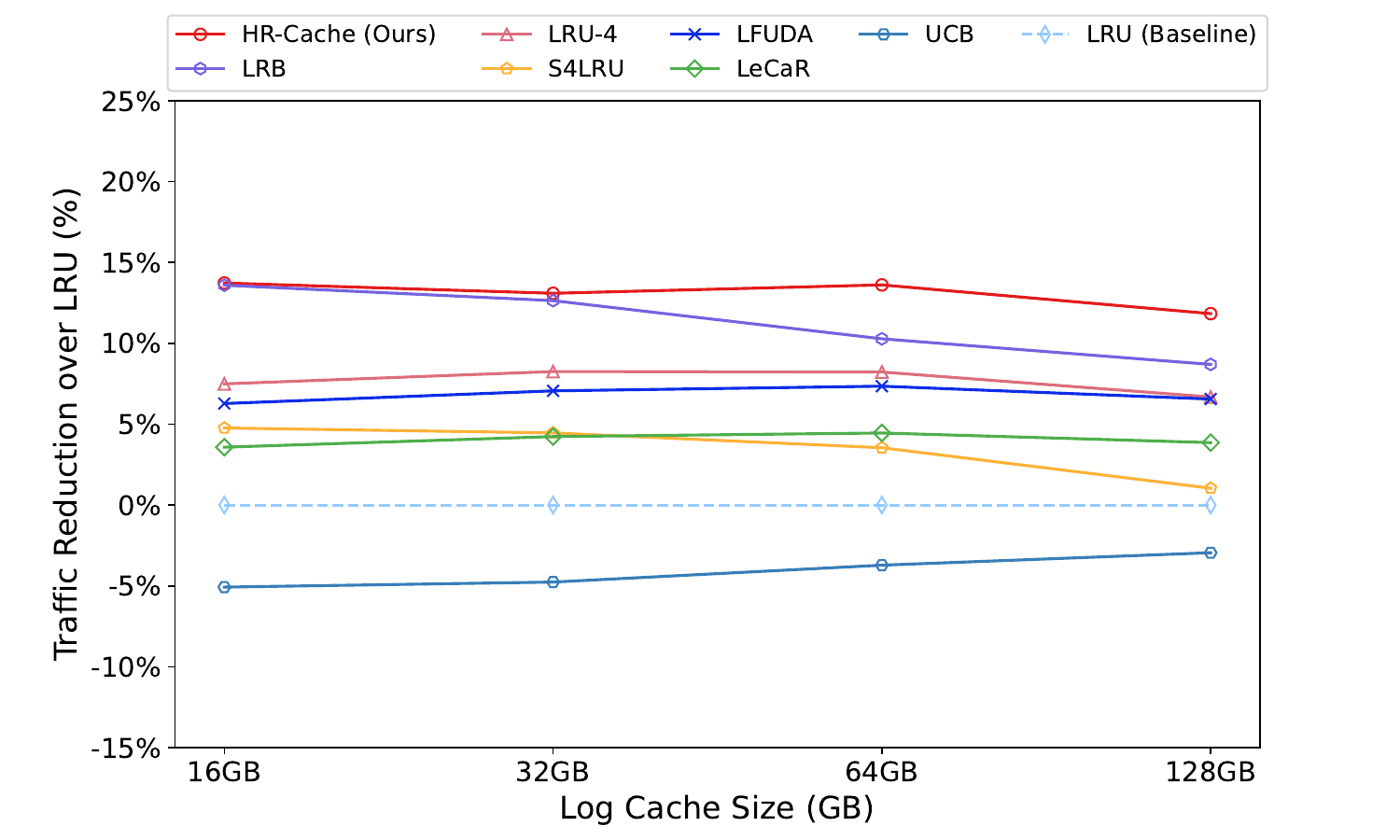}
    \caption{Wikipedia 2019}
    \label{fig:wiki19_traffic}
  \end{subfigure}

  \caption{WAN traffic reduction compared to LRU across various cache sizes for HR-Cache and seven leading algorithms. HR-Cache consistently achieves 2.2–14.6\% greater WAN traffic savings than LRU, outperforming the SOA alternatives.}
  \label{fig:traffic_charts}
\end{figure*}
\subsection{Experimental Methodology}
This subsection describes traces, the experiment setup of our simulation, the competing algorithms, and the parameter settings of HR-Cache. Unless otherwise noted, the reported results for HR-Cache are based on its default operation settings, which include batch-mode inference with a batch size of 128.

\subsubsection*{Workloads}
Our evaluation uses a set of four distinct traces to create a diverse testing environment for the HR-Cache system. This includes two public CDN production traces from Wikipedia for the years 2018 and 2019 \cite{song2020learning}, a public trace from CloudPhysics \cite{waldspurger2015efficient}, and a synthetic trace generated using the JEDI tool in \cite{sabnis2022jedi, sabnis2021tragen}. The selection of these traces aims to represent the performance of HR-Cache across a wide spectrum of real-world and synthetic workloads. Detailed descriptions of each trace source are as follows:

\begin{enumerate}
    \item \textbf{Wikipedia Traces (2018 and 2019)}: These traces are sourced from Content Delivery Network (CDN) nodes in a metropolitan area in 2018 and 2019, respectively. They mainly consist of web and multimedia content, including images and videos, catering to Wikipedia pages. To reflect the typical environment of edge caches, our evaluations on these traces are conducted with cache sizes of 16 GB, 32 GB, 64 GB, and 128 GB, aligning with the characteristics of smaller cache sizes often found in edge caches \cite{liu2016caching}.
    
    \item \textbf{CloudPhysics Trace}: A Block I/O trace from \cite{waldspurger2015efficient}, capturing the activity of VMware virtual disks. This trace introduces a more diverse workload for our analysis, extending beyond the typical CDN scenarios to demonstrate the generalizability of our method across different computing environments. In our analysis of this trace, we chose cache sizes of 1 GB, 4 GB, 8 GB, and 16 GB, reflecting common configurations in virtual machine environments.
    
    \item \textbf{EU Synthetic Trace}: This trace is generated using the JEDI tool \cite{sabnis2022jedi} which produces traces that have similar caching properties and object-level properties as original production traces. We use the ``eu'' traffic class which is tailored to replicate the traffic patterns observed in an Akamai’s production CDN, specifically those serving content related to social media. For this trace, we use cache sizes of 256 GB, 512 GB, 1 TB, and 2 TB. This decision is based on the trace's large working set size, where smaller cache sizes would not be effective or meaningful for performance analysis.
\end{enumerate}
Table ~\ref{table:summary} summarizes key properties of the four traces.

\subsubsection*{State-of-the-art algorithms}\label{sec: soa_alg}
In our evaluation, HR-Cache is compared with twelve state-of-the-art caching algorithms: LRB, LRU, LRU-4, S4LRU, GDSF, LFUDA, AdaptSize, Hyperbolic, LHD, LeCaR, and UCB. To enhance readability, we present only the six best-performing algorithms compared to LRU. These are divided into two categories: 1) learning-based algorithms, which include LRB \cite{song2020learning}, LeCaR \cite{vietri2018driving}, and UCB \cite{costa2017mlcache}; and 2) heuristics-based algorithms, comprising LRU-4 \cite{o1993lru}, LFUDA \cite{arlitt2000evaluating}, and S4LRU\cite{huang2013analysis}. 

\begin{table}[h]
\centering
\begin{tabular}{|l|l|}
\hline
\textbf{Parameter} & \textbf{Value} \\ \hline
Learning Rate          & 0.1            \\ \hline
Max Depth              & 50             \\ \hline
Number of Trees        & 100            \\ \hline
Max Number of Bins     & 255            \\ \hline
Objective              & logistic regression   \\ \hline
\end{tabular}
\caption{Parameters of the GBDT Model}
\label{tab:my_label}
\end{table}
\subsubsection*{Experimental Setup}
All simulation experiments were run on a Google Cloud server with 24 E2-v CPUs (12 shared physical cores) and 64 GB of RAM.
Unless specified otherwise, the reported results for HR-Cache are based on the settings that HR-Cache operates in batch-mode inference with a batch size of 128. We also set the frequency decay factor to 0.9 for the decayed frequency feature. Throughout our evaluation, we utilized the parameters listed in Table \ref{tab:my_label} for the GBDT model in LightGBM.

We note that the LRB algorithm was run using its default window parameter. The longer duration of this default memory window, in comparison to the lengths of our traces and the sizes of our caches, might have a bearing on its performance. However, any such influence is expected to be advantageous, which contributes to a balanced comparison in our study.

In all our experiments, the initial training window, during which HR-Cache reverts to LRU, is considered a warm-up phase. We report the metrics for HR-Cache and other algorithms after this period. Notably, LRB starts its training ahead of our framework, and thus, this warm-up phase provides enough time for its training phase to start.

\subsection{Main Results}
We compare HR-Cache with the caching algorithms detailed in Section~\ref{sec: soa_alg}, utilizing simulations across various cache sizes. Figure~\ref{fig:traffic_charts} illustrates the reduction in wide-area network (WAN) traffic for each algorithm relative to LRU, across different cache sizes and the six traces.

HR-Cache consistently outperforms existing state-of-the-art algorithms, securing the lowest byte miss ratios across various combinations of traces and cache sizes. The sole exception is observed with the EU Synthetic, size 2048, where HR-Cache achieves performance equivalent to that of LeCar. On average, HR-Cache reduces WAN traffic by over 9.7\% compared to LRU, with reductions ranging from 2.2–14.6\%. Its robust performance is evident across all traces, unlike other algorithms that lack consistent improvements across varying traces and cache sizes.

For instance, LRU-4 improves performance over LRU in 3 of the workloads, but completely underperforms in the EU traces, resulting in a significant 16-24\% increase in traffic over LRU (not depicted in the plot due to being below the y-axis). On the other hand, UCB generally underperforms compared to the other algorithms, with a notable exception in CloudPhysics at 16 GB, where it closely rivals HR-Cache and LRB.
Shifting focus to LeCaR and LFUDA, these algorithms consistently outperform LRU, yet they do not manage to surpass the effectiveness of other top-performing policies. LRB, on the other hand, exhibits strong results on the Wiki traces, however, it performs the same or falls short in comparison to HR-Cache even where it performs best. Moreover, LRB is outperformed by heuristic algorithms in an instance of CloudPhysics and EU Synthetic and undergoes a significant decrease in effectiveness in the EU Synthetic trace, particularly as cache sizes increase.

Furthermore, it is important to note that the pattern of WAN traffic reduction achieved by HR-Cache does not consistently correlate with cache capacity. For instance, in the EU Synthetic trace, we observe that the traffic reduction effectively doubles when moving from 256 GB to 1 TB. Conversely, in CloudPhysics, HR-Cache's reduction over LRU generally shows an increasing trend, yet there are instances where the improvement trend inversely declines. This variability suggests that the traces used in our study encompass a diverse array of request patterns, influencing the performance dynamics of HR-Cache differently across scenarios.

Overall, these results suggest that heuristic-based algorithms excel with specific patterns but falter with others. A similar trend is observed among the learning-based algorithms we evaluated. UCB generally underperforms across the board, and LeCaR struggles to match the performance of state-of-the-art alternatives. LRB, although demonstrating strengths in certain scenarios, does not consistently show improvement, underscoring the variability in its efficacy.
\begin{figure}[!ht]
  \centering
  \begin{subfigure}[b]{0.5\columnwidth}
    \includegraphics[width=\linewidth]{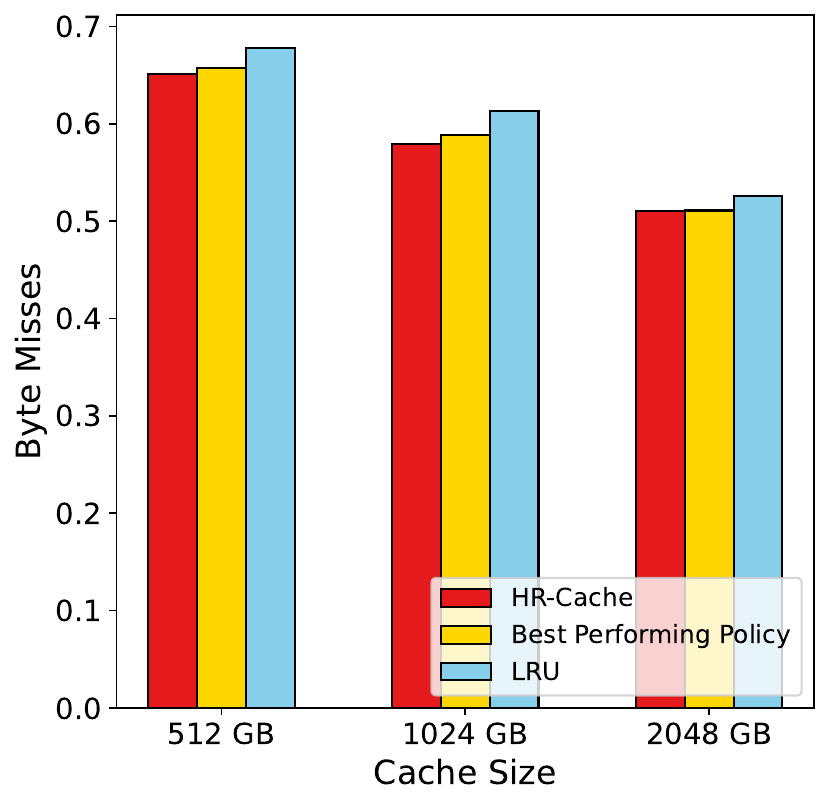}
    \caption{EU Synthetic}
    \label{fig:sub1}
  \end{subfigure}%
  \begin{subfigure}[b]{0.5\columnwidth}
    \includegraphics[width=\linewidth]{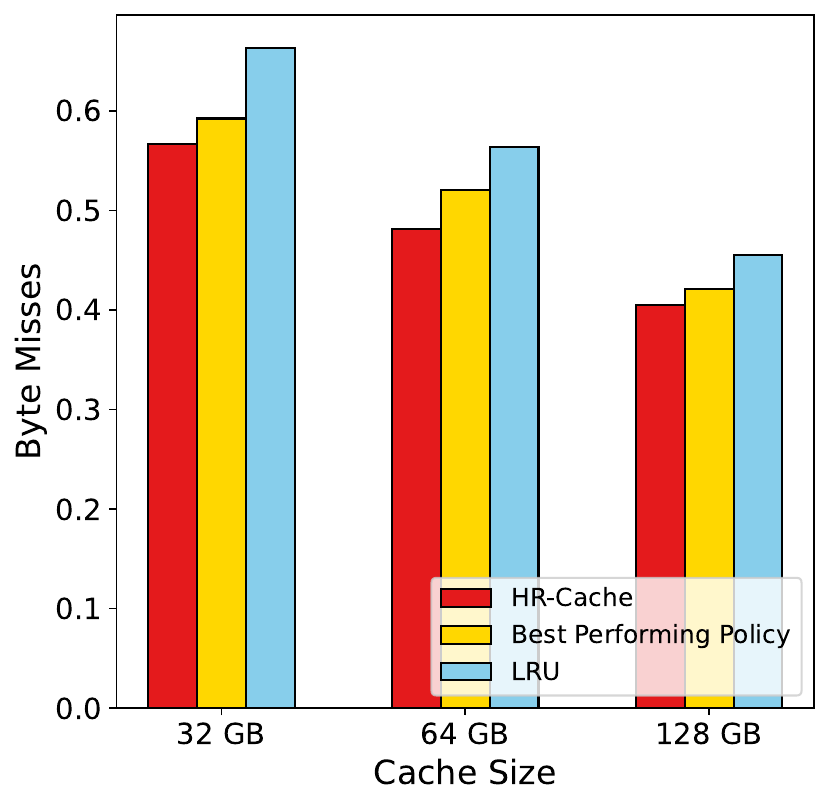}
    \caption{Wiki 2018}
    \label{fig:sub2}
  \end{subfigure}
  \begin{subfigure}[b]{0.5\columnwidth}
    \includegraphics[width=\linewidth]{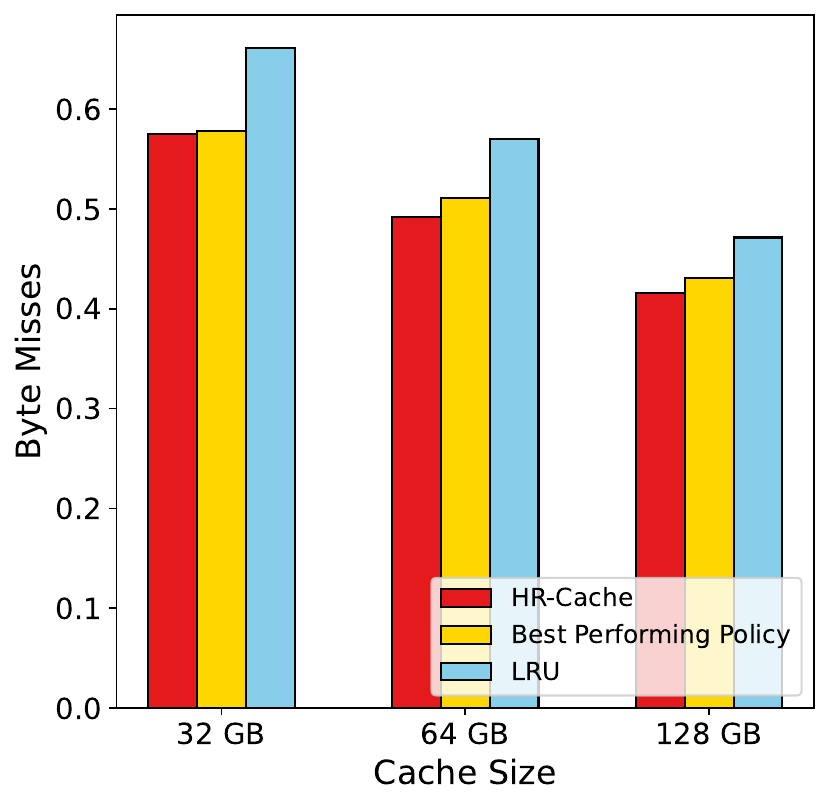}
    \caption{Wiki 2019}
    \label{fig:sub3}
  \end{subfigure}%
  \begin{subfigure}[b]{0.5\columnwidth}
    \includegraphics[width=\linewidth]{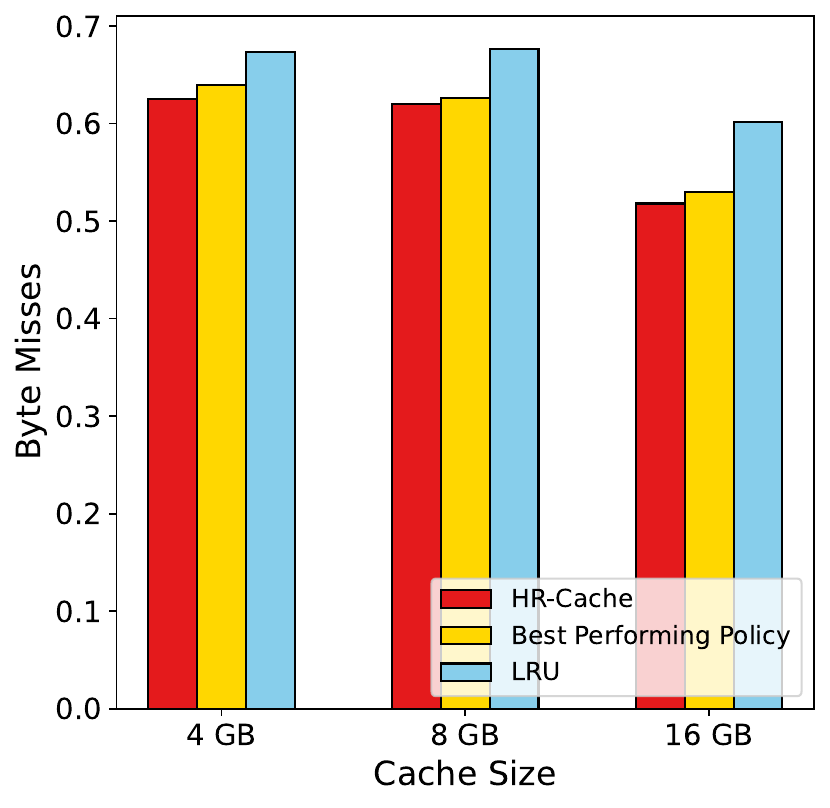}
    \caption{CloudPhysics}
    \label{fig:sub4}
  \end{subfigure}
  \caption{Comparison of Byte Miss Ratios for HR-Cache, the Best Performing Policy, and LRU}
  \label{fig:main}
\end{figure}

\subsection{Prediction Overhead Optimization} \label{sec:optimization}
In this section, we analyze the additional prediction overhead introduced by HR-Cache in comparison with the state-of-the-art LRB algorithm. To understand the source of this overhead in both LRB and HR-Cache, we examine this overhead for the Wiki 2018 trace.

LRB incurs prediction overhead by running predictions on 64 samples for each eviction event. In contrast, HR-Cache requires a prediction for each incoming request to determine cache-friendliness. However, HR-Cache's batch mode significantly reduces this requirement by enabling inference on every 128 requests, rather than on each individual request.
As both frameworks utilize the GBDT model, we measure the inference time for batches of 64 (LRB's eviction candidate count) and 128 (HR-Cache's inference batch size) inputs, respectively. The results of these measurements is found in Table~\ref{tab:pred_overhead}. 

\begin{table}[ht]
\centering
\captionsetup{font=small} 
\small
\begin{tabular}{|c|c|c|}
\hline
{} & Prediction batch & Prediction time ($\mu s$) \\ \hline
LRB & 64 & 183 \\ \hline
HR-Cache & 128 & 220 \\ \hline
\end{tabular}
\vspace{5pt} 
\caption{Comparison of Prediction Batch Sizes and Prediction Times for LRB and HR-Cache Algorithms}
\label{tab:pred_overhead}
\end{table} 
\vspace{-5pt} 

Given that LRB is required to run predictions with every eviction event, its prediction overhead is directly tied to the object miss ratio. For our analysis, we assume the best-case scenario for LRB, where only one object needs to be evicted per cache miss. 

Under the Wikipedia 2018 workload for cache sizes of 64 GB and 128 GB, LRB is required to run predictions for 18\% and 13\% of requests, respectively. In contrast, HR-Cache with a batch size of 128, effectively runs predictions for only $1/128$ of requests. Taking this and the measured inference times into account, this translates to a prediction overhead reduction by factors of 19.2x and 13.8x for cache sizes of 64 GB and 128 GB, respectively, when compared to LRB.

\begin{table}[ht]
\centering
\caption{Prediction Overhead Reduction For Wiki 2018}
\label{tab:algorithm_comparison}
\small
\begin{tabular}{|l|cc|cc|cc|}
\hline
\multirow{2}{*}{Alg.} & \multicolumn{2}{c|}{Miss Ratio} & \multicolumn{2}{c|}{Pred Time ($\mu s$/$req$)} & \multicolumn{2}{c|}{Reduction Factor} \\
                      & 64 GB  & 128 GB & 64 GB & 128 GB & 64 GB & 128 GB \\ \hline
LRB                   & 0.18   & 0.13  & 32.94  & 23.8    & -     & -     \\
HR-Cache              & -   & -  & 1.72  & 1.72    & 19.2x   & 13.8x   \\ \hline
\end{tabular}
\end{table}

Another aspect of overhead comes from the process of feature building. HR-Cache constructs one feature per request, while LRB, in contrast, needs to build 64 features on each object miss. This difference results in a significant reduction of overhead for HR-Cache. Specifically, under the Wiki 2018 workload for cache sizes of 64 GB and 128 GB, HR-Cache achieves a reduction in feature-building overhead by factors of 11.5x and 8.3x, respectively, compared to LRB.

To illustrate HR-Cache's computational burden, consider the Wiki 2018 trace with a cache size of 64: replaying 84 million requests, conducting frequent training and inference, and updating our cache based on these predictions, takes approximately 13 minutes, which is more than acceptable given the inter-arrival request rates for objects.

\subsection{Ablation}
\begin{figure}[ht]
  \centering
  \includegraphics[width=\linewidth]{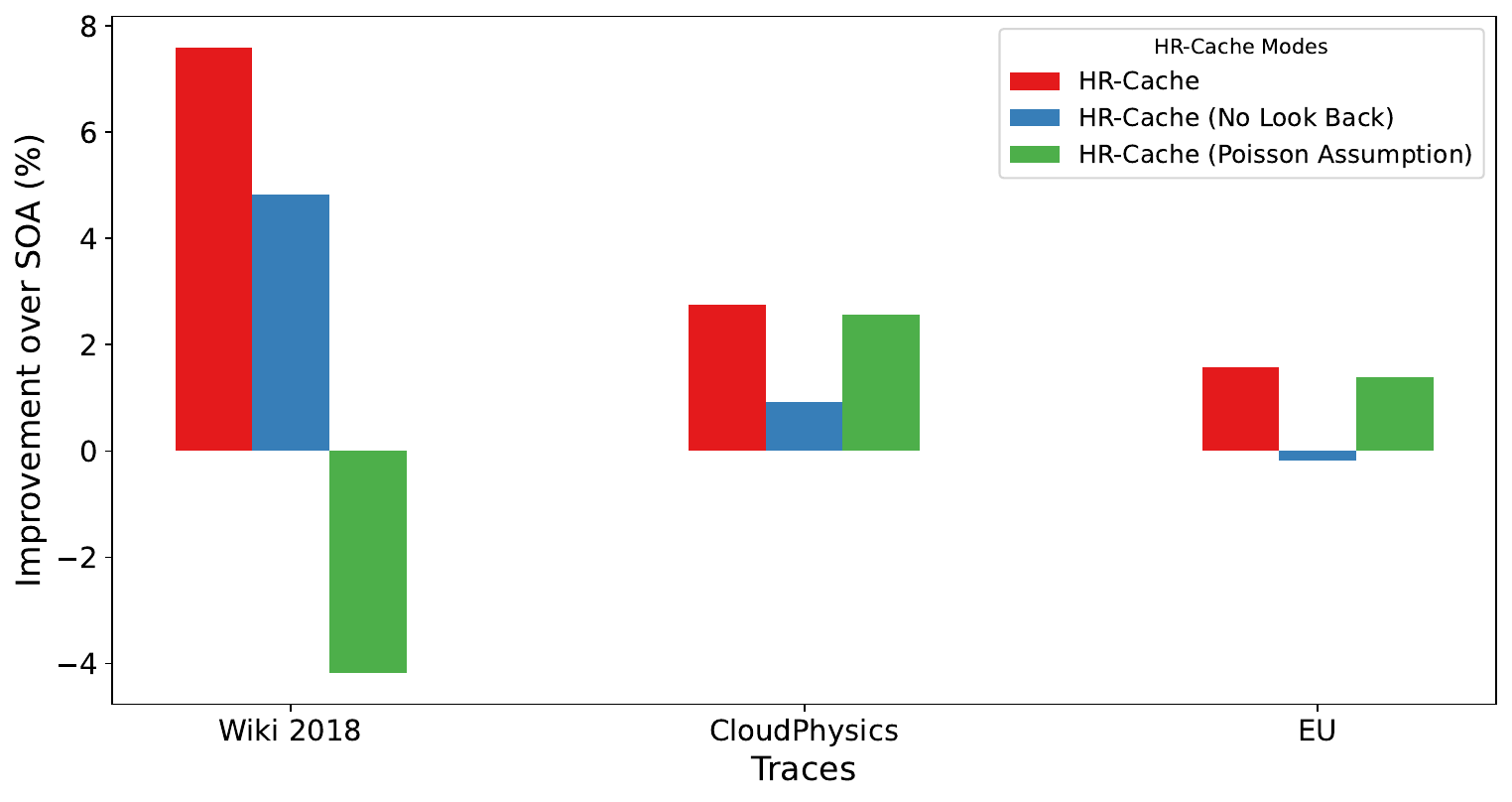}
  \caption{Percentage improvement of HR-Cache modes over the state-of-the-art method.}
  \label{fig:hr-cache-ablation}
\end{figure}

In Section \ref{sec: derive_hrrule}, we discussed how hit or miss outcomes determined by hazard ordering may not directly correspond to cache decisions. This is because Hazard Rate Ordering (HRO) assumes objects with the highest hazard rates are always pre-fetched and available in the cache whenever a request occurs at time $t$. Therefore, if a request for an object at time $t$ is a hit, we previously classify it as cache-friendly in its last request.
We also noted that modeling the request process as a Poisson process is a simplification, even though it offers a less complex method for calculating hazard rates. Under this process, the hazard rate remains constant. In our study, we conduct an ablation analysis on three of the traces, where we remove the look-back option in one scenario. In another, we operate HR-Cache under the Poisson assumption, as opposed to using kernel hazard estimation. Figure ~\ref{fig:hr-cache-ablation} shows the relative gains of our assumptions. 
On the Wiki 2018 trace, the look-back option assumption significantly influence performance, a trend also observed in the CloudPhysics and EU traces. Under the Poisson assumption, the performance on the Wiki 2018 trace is markedly diminished, whereas this assumption has a minimal impact on the other two traces. This disparity also confirms our hypothesis that the Poisson assumption may not be universally applicable to all real-world traces.

\section{CONCLUSION AND FUTURE WORK} \label{conclusion}
In this paper, we proposed HR-Cache, a novel learning-based caching framework for edge environments. It learns from hazard rate ordering decisions to identify cache-averse objects and prioritizes them for eviction. HR-Cache comprises two main components: it reconstructs the hazard rate ordering on a window of requests using kernel hazard estimation and a decision tree classifier that learns to predict the cache-friendliness of incoming requests. We evaluated our framework using real-world data traces and compared it with several state-of-the-art caching strategies. Our results indicate that HR-Cache significantly improves the byte hit rate compared to LRU and surpasses a wide range of state-of-the-art policies, all while maintaining minimal prediction overhead compared to contemporary learning-based cache policies. Further experiments were conducted to confirm the positive impact of our specific design choices and assumptions on HR-Cache's performance, highlighting their validity and effectiveness.

We envision two main ways for future exploration to enhance HR-Cache applicability and performance in real-world scenarios. The first involves extending HR-Cache for distributed caching environments. Exploring the potential of HR-Cache in distributed environments opens avenues for leveraging federated learning to pool insights from diverse data sources, enhancing model accuracy while adhering to privacy concerns. This approach, however, introduces challenges, particularly with non-IID data, which could affect model performance. Additionally, adapting HR-Cache to hierarchical cache architectures allows us to optimize cache utilization across different levels and closely mirrors the operational structures of real-world CDN and edge caches. This adaptation, however, necessitates navigating the intricacies of cache dynamics within such structures.
The second avenue focuses on integrating HR-Cache into production cache systems, necessitating adjustments to overcome hardware limitations and ensure seamless operation. This effort will extend to evaluating HR-Cache's impact on latency, resource consumption, and scalability under practical conditions.

\section*{Acknowledgment}

This research was supported by the Natural Sciences and Engineering Research Council of Canada (NSERC) through the NSERC CREATE grant, ``Dependable Internet of Things Applications (DITA).''

\bibliographystyle{ACM-Reference-Format}
\balance
\bibliography{ref}

\end{document}